\documentstyle[12pt]{ets2000}
\def\rfr#1{eq.(\ref{#1})}

\def\Rfr#1{Eq.(\ref{#1})}

\def\eqi{\begin{equation}}
\def\eqf{\end{equation}}
\def\eqia{\begin{eqnarray}}
\def\eqfa{\end{eqnarray}}
\def\lb#1{\label{#1}}
\def\rp#1#2{{#1\over#2}}
\def\ci{\cos{i}}
\begin{document}
\title{
Testing General Relativity with LAGEOS, LAGEOS II and Ajisai laser-ranged
satellites}
\author{
Lorenzo Iorio$^{1)}$
\thanks{Fax: +390805443144, E-mail: Lorenzo.Iorio@ba.infn.it }}
\affil{
1) Dipartimento Interateneo di Fisica, \\
Via Amendola 173, 70126, Italy}
\date{
(Received May, 2001; Revised; Accepted)}
\footnotetext{ Fax: +390805443144, E-mail:
Lorenzo.Iorio@ba.infn.it }
\begin{abstract}
The accuracy reached in the past few years by Satellite Laser
Ranging (SLR) allows for measuring even tiny features of the
Earth's gravitational field predicted by Einstein's General
Relativity by means of artificial satellites. The gravitomagnetic
dragging of the orbit of a test body is currently under
measurement by analyzing a suitable combination of the orbital
residuals of LAGEOS and LAGEOS II. The lower bound of the error in
this experiment amount to $12.92\%$. It is due  to the mismodeling
in the even zonal harmonics of the geopotential which are the most
important sources of systematic error. A similar approach could be
used in order to measure the relativistic gravitoelectric
pericenter shift in the field of the Earth with a lower bound of
the systematic relative error of $6.59\cdot 10^{-3}$ due to the
even zonal harmonics as well. The inclusion of the ranging data to
the Japanese passive geodetic satellite Ajisai would improve such
limits to $10.78\%$ and $8.1\cdot 10^{-4}$ respectively and would allow to improve
the accuracy in the determination of the PPN parameters $\beta$ and $\gamma$.
\end{abstract}
\section{Introduction}
General Relativity, in its slow-motion and weak-field
approximation, predicts that a central spherically symmetric body,
both if it rotates and if it is static, induces on the Keplerian
orbital elements (Sterne, 1960) of a test body orbiting it certain
small effects which are unknown in Newtonian classical mechanics
(Ciufolini and Wheeler, 1995).

The most famous of them is the well known gravitoelectric
precession of the pericenter $\omega$ generated by the
Schwarzschild's metric of a central, static spherical body
(Ciufolini and Wheeler, 1995).
It amounts to: \eqi \dot\omega_{\rm GR}=\rp{3 n G M}{c^2 a
(1-e^2)},\lb{peg}\eqf in which $G$ is the Newtonian gravitational
constant, $c$ is the speed of light in vacuum, $M$ is the mass
of the central object, $a$ and $e$ are the semimajor axis and the
eccentricity, respectively, of the orbit of the test body and
$n=\sqrt{GM/a^{3}}$ is its mean motion. Such effect was detected,
up to now, by measuring with the radar ranging technique the
Mercury's perihelion advance in the field of the Sun at an
accuracy level of the order of $1\%$ by including
certain sources of systematic errors (Shapiro et al., 1972;
Shapiro, 1990). It constitutes one of the classical tests of
General Relativity; unfortunately, its interpretation
is affected by the currently still existing uncertainty
in the quadrupole mass moment $J_{\odot}$ of the Sun 
(Ciufolini and Wheeler, 1995; Pireaux et al., 2001). 
A first attempt to measure it in the field of
the Earth by analyzing the laser ranging data to LAGEOS SLR
satellite was reported in (Ciufolini and Matzner, 1992), but the
accuracy was $20\%$ only. For LAGEOS and LAGEOS II the
relativistic gravitoelectric perigee precession  due to the
Earth's mass amount to 3312.35 and 3387.46 milliarcseconds per
year ({\rm mas/y} in the following) respectively.

Much more smaller is the effect of the proper angular momentum $J$
of the central spherical body on the node $\Omega$ and the perigee
$\omega$ of the orbiting test body. Such feature, derived from the
Einstein's equations for the first time by Lense and Thirring in
1918 (Lense and Thirring, 1918), is called gravitomagnetism
because the general relativistic linearized equations of motion of
a test body in the gravitational field of a central rotating body
are formally analogous to that governing the motion of an
electrically charged particle acted upon by electric and magnetic
fields. The gravitomagnetic rates of the node and the perigee of a
test body are:
\begin{eqnarray}
\dot\Omega_{\rm LT}&=&\rp{2GJ}{c^{2}a^{3}(1-e^2)^{3/2}},\\
\dot\omega_{\rm LT}&=&-\rp{6GJ\ci}{c^{2}a^{3}(1-e^2)^{3/2}},
\end{eqnarray}
where $i$ is the inclination of the orbital plane to the
equatorial plane of the central body. The Lense-Thirring effect
was measured for the first time in the field of the Earth by
Ciufolini and coworkers in 1998 (Ciufolini et al., 1998;
Ciufolini, 2000) with an accuracy of $20\%$ over a time span of 4
years. They analyzed the orbits of LAGEOS and LAGEOS II. For them
the gravitomagnetic precessions induced by the rotation of the
Earth, whose angular momentum is $J_{\oplus}=5.9\times 10^{33}$
kg m$^{2}$ s$^{-1}$, amount to:
\begin{eqnarray} \dot
\Omega_{\rm LT}^{\rm LAGEOS}&\simeq& 31 \ {\rm mas/y},\\ \dot
\Omega_{\rm LT}^{\rm LAGEOS II}&\simeq& 31.5 \ {\rm mas/y}, \\
\dot \omega_{\rm LT}^{\rm LAGEOS}&\simeq& 31.6\ {\rm mas/y},\\
\dot\omega_{\rm LT}^{\rm LAGEOS II}&\simeq& -57\ {\rm mas/y}.
 \end{eqnarray}
Here we want to explore the possibility of refining the precision
of the measurements of such general relativistic effects in the
terrestrial field  by using the laser ranged data to Ajisai as
well.

The paper is organized as follows: In Section 2 we explain the
role of the error budget in such kinds of experiment pointing out
what is the major source of systematic uncertainty. In Section 3
we show how the use of  data from Ajisai could improve the
accuracy of the measurements of the gravitomagnetic Lense-Thirring
effect and of the gravitoelectric perigee shift. Section 4 is
devoted to the conclusions.

In Table 1 we summarize the values of the orbital parameters of
LAGEOS, LAGEOS II and Ajisai.
\begin{table}[h]
\caption{Orbital parameters of LAGEOS and LAGEOS II and Ajisai satellites.}
\begin{center}
\begin{tabular}{lllll}
\noalign{\hrule height 1.5pt} Orbital Parameter & LAGEOS & LAGEOS
II & Ajisai \\ \hline \textit{a} & $12.27\times 10^{6}$ m &
$12.163 \times
10^{6}$ m & $7.87\times 10^{6}$ m \\
\textit{e} & 0.0045 & 0.014 & 0.001 \\
\textit{i} & 110 deg & 52.65 deg & 50 deg \\
\textit{n} & $4.65\times 10^{-4}\ {\rm s}^{-1}$ & $4.71\times
10^{-4}\ {\rm s}^{-1}$ & $9.05\times 10^{-4}\ {\rm s}^{-1}$ \\
\noalign{\hrule
height 1.5pt}
\end{tabular}
\end{center}
\end{table}

\section{Error analysis}

When a satellite-based experiment is planned in order to detect
such relativistic effects on the orbital elements of artificial
satellites in the gravitational field of the Earth the correct
evaluation of the error budget is of the utmost importance.
Indeed, the accuracy of these measurements is affected mainly by
the systematic errors induced by the mismodeling in the various
competing forces of the terrestrial environment (Montenbruck and
Gill, 2000) which in many cases are quite larger than the
relativistic features of interest. For example, in the case of
LAGEOS and LAGEOS II, if we analyzed the node or the perigee of a single
satellite the error induced on the classical rates of these
orbital elements by the bad knowledge of the first two even zonal
harmonics $J_2,\ J_4$ of the geopotential (Kaula, 1966) would
affect sensibly the accuracy of the measurement of the
relativistic rates. Regarding the gravitomagnetic shift, such kind
of errors would be even larger than the relativistic effect itself
for the orbital element considered (Ciufolini, 1996).

In order to reduce the impact of the mismodeled even zonal
harmonics of the geopotential on the measurement of the
Lense-Thirring effect in 1996 Ciufolini (1996) put forward an
interesting strategy based on the use of a suitably weighted
combination of the orbital residuals of the nodes of LAGEOS and
LAGEOS II and the perigee of LAGEOS II. It is:
\eqi\delta\dot\Omega^{\rm I}+c_1
\times\delta\dot\Omega^{\rm II}+c_2\times\delta\dot\omega^{\rm II}=\mu_{\rm LT}
\times 60.2.\lb{unolt}\eqf In it $c_1=0.295$, $c_2=-0.35$,
$\mu_{\rm LT}$ is the scaling parameter{\footnotetext{$\mu_{\rm LT}$ can be expressed
in terms of the PPN parameter $\gamma$ as $\rp{(1+\gamma)}{2}$,
but this fact is not particularly relevant since the relatively large uncertainty in the measurement of $\mu_{\rm LT}$
makes it unsuitable in order to constraint seriously $\gamma$. In order
to meet this requirement 
other experiments have been proven more useful (Will, 1993).}}, equal to 1 in General
Relativity and 0 in classical mechanics, to be determined with
least-squares fits and $\delta\Omega^{\rm I},\ \delta\Omega^{\rm II},\
\delta\omega^{\rm II}$ are the orbital residuals, in mas, calculated
with the aid of some orbit determination software like UTOPIA or
GEODYN, of the nodes of LAGEOS and LAGEOS II and the perigee of
LAGEOS II. General Relativity predicts for \rfr{unolt} a linear
trend with a slope of 60.2 {\rm mas/y}. {\footnotetext{Since the
observable for the perigee is $ea\dot\omega$ and
$e_{\rm LAGEOS}=0.0045$, the perigee of LAGEOS turns out to be
difficult to be measured accurately enough to detect its
gravitomagnetic shift.}} \Rfr{unolt} is obtained writing down 
the equations of the residuals of the precessions of the nodes of LAGEOS and LAGEOS II and the 
perigee of LAGEOS II as a 
non homogeneous linear algebraic system
of three equations in the three unknowns $\delta J_2,\ \delta J_4$ and $\mu_{\rm LT}$ and solving for $\mu_{\rm LT}$,
so to cancel out
the static and dynamical contributions of the first two even zonal
harmonics of the terrestrial field (Ciufolini, 1996). The so obtained
coefficients $c_1$ and $c_2$
depend only on the orbital parameters of LAGEOS and LAGEOS II. 
However, the other higher
degree even zonal harmonics $J_6,\ J_8,...$ do affect the combined
residuals. They induce an aliasing linear trend which cannot be
removed from the data: one can only assess as more reliably as
possible the systematic error induced by it on the measurement of
$\mu_{\rm LT}$. According to the covariance matrix of the geopotential
coefficients released by the most recent Earth's gravity model
EGM96 (Lemoine et al., 1998), it amounts to
$\delta\mu_{zonals}=12.92\%\mu_{\rm LT}$. It is important to stress
that it represents the  lower bound of the total systematic error.

A similar strategy could be followed for a new, more precise
measurement of the gravitoelectric perigee shift in the field of
the Earth by means of the residual combination (Iorio et al., 2001): \eqi\delta\dot\omega^{\rm II}+k_1
\times\delta\dot\Omega^{\rm II}+k_2\times\delta\dot\Omega^{\rm I}=\nu_{\rm GR}
\times 3387.46.\lb{grel}\eqf In it $k_1=-0.868$, $k_2=-2.855$,
$\nu_{\rm GR}=(2+2\gamma-\beta)/3$ is the scaling parameter, equal to 1 in General
Relativity and 0 in classical mechanics, to be determined with
least-squares fits as well. It is built up with the PPN parameters
$\beta$ and $\gamma$ in terms of which the alternative metric theories of
gravitation are usually expressed (Will, 1993): 
in General Relativity $\beta=\gamma=1$. General Relativity predicts for
\rfr{grel} a linear trend with a slope of 3387.46 {\rm mas/y}, entirely
due to the gravitoelectric shift of LAGEOS II perigee. Also this
combination allows to cancel out the effects of $J_2$ and $J_4$.
In this case the error induced by the remaining mismodeled zonal
harmonics of the static part of the geopotential amounts to
$\delta\nu_{zonals}=0.6\%\nu_{\rm GR}$.

The other sources of systematic errors are represented by long
period harmonic mismodeled perturbations like tides, solar
radiation pressure, mismodeling in the satellites' inclinations
$i$, etc. (Ciufolini et al., 1997). Their impact can be reduced by
using adequately long time spans $T_{obs}$ or, if their periods
$P$ are shorter than $T_{obs}$, they can be viewed as empirically
fitted quantities and removed from the signal. Moreover, if the
time span is an integer multiple of their periods they average
out. Their effect on the combined residuals is reduced if the
coefficients entering the combinations are smaller than unity, as
it is in these cases. For a recent analysis of their impact on the
gravitomagnetic LAGEOS experiment see (Iorio and Pavlis, 2001).

\section{The role of Ajisai}
For a given residual combination, the systematic error induced by
aliasing linear trends, like that produced by the mismodeled
static part of the geopotential, neither reduces as the time goes
along nor can be eliminated by handling suitably the data. A way
to reduce such error could be the inclusion in the residual
combinations of more orbital elements so to cancel out other
higher degree even zonal harmonics $J_{2n},\ n\geq 3$. If
possible, the choice of the additional orbital elements should be
done in order to enhance the slope of the relativistic trends;
moreover, they should not introduce too large additional
perturbations. This goal restricts our choice to the node and the
perigee of the other existing passive geodetic laser-tracked
satellites Ajisai, Starlette, Stella, Westpac-1, Etalon-1 and
Etalon-2. For a previous analysis of their possible use  in
measuring the gravitomagnetic effect see (Casotto et al., 1990).
The perigee is a very "dirty" element due to the large number of
gravitational and nongravitational perturbations affecting it and
it could be measured with low accuracy because the orbits of most
of the geodetic satellites are even less elliptical than that of
LAGEOS and LAGEOS II. So, we focus our attention on the node which is affected
by the gravitomagnetic force (but not by the relativistic
gravitoelectric  one), is one of the most accurately measurable
orbital elements and is less sensitive to the orbital
perturbations of the terrestrial environment than the perigee.

It turns out that the contributions of the mismodeled higher
degree even zonal harmonics to the classical nodal rates of
Starlette, Stella and Westpac-1, due to their lower altitude, are
too large and would raise the error in the relativistic trends.
About the Lense-Thirring effect, in Table 2 we report the most favourable alternative
combinations including the perigee of LAGEOS II
and in Table 3 the related systematic errors due to the remaining
zonal harmonics are quoted. It is assumed that $\delta\dot\Omega^{\rm I}$ is always present
multiplied by 1. The slopes of the gravitomagnetic trends, in mas/y, are denoted by $x_{\rm LT}$ and
$\delta\mu_{\rm LT}$ represents the percent systematic error due to the even zonal harmonics of the geopotential. 
The combination by Ciufolini is denoted by C. In Table 2 Aji=Ajisai, Str=Starlette, Stl=Stella and WS=Westpac-1.

\begin{table}[h]
\caption{Alternative combinations}
\begin{center}
\begin{tabular}{lllllll}
\noalign{\hrule height 1.5pt}  & $\Omega^{\rm II}$ & $\Omega^{\rm Aji}$
 & $\Omega^{\rm Str}$ & $\Omega^{\rm Stl}$ & $\Omega^{\rm WS}$ & $\omega^{\rm II}$ \\ 
\hline
C & x & 0 & 0 & 0 & 0 & x \\
1 & x & x & 0 & 0 & 0 & x \\
2 & x & x & 0 & x & 0 & x \\
3 & x & x & x & x & 0 & x \\
4 & x & 0 & x & x & 0 & x \\
5 & x & 0 & x & 0 & x & x \\
6 & x & x & 0 & 0 & x & x \\
7 & x & x & x & 0 & x & x \\
8 & x & 0 & x & 0 & 0 & x \\
\noalign{\hrule
height 1.5pt}
\end{tabular}
\end{center}
\end{table}


\begin{table}[h]
\caption{Alternative combinations: numerical values}
\begin{center}
\begin{tabular}{llllllll}
\noalign{\hrule height 1.5pt}  & $c_1$ & $c_2$ & $c_3$ & $c_4$ & $c_5$ & $x_{\rm LT}$ (mas/y) & $\delta\mu_{\rm LT}\ (\%)$ \\ 
\hline 
C & 0.295 & -0.35 & 0 & 0 & 0 & 60.2 & 12.92 \\ 
1 & 0.443 & -0.0275 & -0.341  & 0 & 0 & 61.26 & 10.78 \\ 
2 & 0.405 & -0.017  & -0.0087 & -0.306 & 0 & 57.85 & 13.6 \\ 
3 & 0.407 & -0.0109 & 0.001   & -0.0088 & -0.305 & 57.85 & 13.65 \\ 
4 & 0.38  & -0.0096 & -0.0085 & -0.307 & 0 & 57.85 & 14.75 \\
5 & 0.381 & -0.0097 & -0.00904 & -0.308 & 0 & 57.82 & 16.46 \\
6 & 0.405 & -0.017  & -0.0093  & -0.306 & 0 & 57.82 & 16.5 \\
7 & 0.407 & -0.018  & 0.00064  & -0.0093 & -0.306 & 57.81 & 16.5 \\ 
8 & 0.404 & -0.015  & -0.343   & 0 & 0 & 61.11 & 16.74 \\
\noalign{\hrule
height 1.5pt}
\end{tabular}
\end{center}
\end{table}


From Table 3 it can be noticed that, with the exception of combination 1, 
all the combinations other than that by Ciufolini
would be not competitive with it because they would be affected by larger systematic errors.   
Etalon-1 and Etalon-2 turn out to be unsuitable because, if included in the
combinations, from one hand they would greatly reduce the error of
the static part of the geopotential,  but from the other hand they
would induce very long period  nonzonal perturbations of tidal
origin which would affect the combined residuals and resemble
linear trends over time spans of few years. Ajisai, in this
optics, lies at an intermediate stage. Its gravitomagnetic nodal
precession amounts to 115.6 {\rm mas/y}. By using its node in order to
cancel out the effect of $J_6$ we obtain for the gravitomagnetic
experiment the combination 1: \eqi\delta\dot\Omega^{\rm I}+c_1
\times\delta\dot\Omega^{\rm II}+c_2\times\delta\dot\Omega^{Aj}
+c_3\times\delta\dot\omega^{\rm II}=\mu_{\rm LT} \times
61.26.\lb{ltaj}\eqf In it $c_1=0.443$, $c_2=-0.0275$ and
$c_3=-0.341$. Regarding the gravitoelectric experiment we have:
\eqi\delta\dot\omega^{\rm II}+k_1
\times\delta\dot\Omega^{\rm II}+k_2\times\delta\dot\Omega^{\rm I}+
k_3\times\delta\dot\Omega^{Aj}+k_4\times\delta\dot\omega^{\rm I}=\nu_{\rm GR}
\times 7928.21.\lb{grelaj}\eqf In it $k_1=-1.962$, $k_2=-3.693$,
$k_3=0.036$  and $k_4=1.370$. The coefficients of Ajisai are smaller than that of
LAGEOS and LAGEOS II due to its different orbital parameters. The inclusion of the residuals of
Ajisai's  node, in fact, would introduce an improvement in
reducing the systematic errors due to the geopotential: indeed,
according to EGM96 model,  they amount to
$\delta\mu_{zonals}=10.78\%\mu_{\rm LT}$ and
$\delta\nu_{zonals}=0.08\%\nu_{\rm GR}$. Furthermore, the coefficients
with which it would enter the modified combinations are
sufficiently small to depress the impact of the other
gravitational and nongravitational perturbations (Sengoku et al.,
1995; 1997) acting on it. This is especially true for the
Lense-Thirring effect: indeed, in \rfr{ltaj} the coefficients of
the elements of
LAGEOS and LAGEOS II are close to those of \rfr{unolt}, while the
coefficient of Ajisai is of the order of $10^{-2}$ only; in
particular, the perigee of LAGEOS II, which is the major source of
mismodeled perturbations, is weighted at the same level. This
means that to the slight improvement in the error due to the
geopotential it should not correspond a worsening of the
time-varying part of the error budget which should be remain
almost the same as in the current LAGEOS experiment.
\section{Conclusions}
The use of suitable combinations of the orbital residuals of the
nodes of LAGEOS, LAGEOS II and Ajisai and the perigee of LAGEOS II
would improve the precision of the measurements of certain subtle
general relativistic effects in the gravitational field of the
Earth. In particular, regarding the Lense-Thirring experiment,
currently performed by analyzing the data of the two LAGEOS and LAGEOS II only,
the systematic error induced by the even zonal harmonics of the
geopotential, which is the major source of uncertainty, would
reduce from the present $12.92\%$ to $10.78\%$ with Ajisai's node.
Concerning the proposed measurement of the relativistic
gravitoelectric pericenter shift, the error induced by the
geopotential would pass from $0.6\%$ to $0.08\%$ by including also
the perigee of LAGEOS and the node of Ajisai.

Regarding the improvements which could be obtained in the accuracy of
our knowledge of the PPN parameters $\beta$ and $\gamma$, if from one hand even a refined version of the 
Lense-Thirring
experiment would not be particularly useful in constraining effectively $\gamma$ through
the measurement of $\mu_{\rm LT}$, 
from the other hand the proposed gravitoelectric experiment (Iorio et al., 2001) 
could be able to obtain interesting results raising
the accuracy on $\gamma$ and $\beta$ to the $10^{-3}-10^{-4}$ level and providing us with an independent
measurement of them in the field of the Earth with laser-ranging.
 
It must be pointed out that both such estimates will greatly
improve in the near future when the new data on the terrestrial
gravitational field will be released by the CHAMP and GRACE
missions.

Concerning
the time-dependent part of the error budget, a detailed analysis
of the impact of the harmonic perturbations of gravitational and
nongravitational origin on the node of Ajisai in the  context of
such relativistic measurements would be required.

\acknowledgments{ I wish to warmly thank F. Vespe for his
important suggestions and help, L. Guerriero for his support to me
at Bari, A. Sengoku for the useful material sent to me, I.
Ciufolini and E. C. Pavlis for their useful observations. }

\begin{thebibliography}{99}
\bibitem{}
Casotto, S., I. Ciufolini, F. Vespe, and G. Bianco (1990): Earth
satellites and Gravitomagnetic Field, Il Nuovo Cimento, {\bf
105B}(5) 589-599.
\bibitem{}
Ciufolini, I. and R. Matzner (1992): Non-Riemannian theories of
gravity and lunar and satellite laser ranging, Int. J. of Mod.
Phys. A, \textbf{7}(4), 843-852.
\bibitem{} Ciufolini, I. and J. A. Wheeler (1995): Gravitation
and Inertia, Princeton University Press, 498p.
\bibitem{}
Ciufolini, I. (1996): On a new method to measure the
gravitomagnetic field using two orbiting satellites, Il Nuovo
Cimento, {\bf 109A},(12), 1709-1720.
\bibitem{}
Ciufolini,
 I., F. Chieppa, D. Lucchesi and  F. Vespe (1997): Test of Lense-Thirring
orbital shift due to spin, Class. Quantum Grav., \textbf{14},
2701-2726.
\bibitem{}
Ciufolini, I., E. Pavlis, F. Chieppa, E. Fernandes-Vieira and J.
P{\'{e}}rez-Mercader (1998): Test of General Relativity and
Measurement of the Lense-Thirring Effect with Two Earth
Satellites, Science, \textbf{279}, 2100-2103.
\bibitem{}
Ciufolini, I. (2000): The 1995-99 measurements of the
Lense-Thirring effect using laser-ranged satellites, Class.
Quantum Grav., \textbf{17}(12), 2369-2380.
\bibitem{} 
Iorio, L., I. Ciufolini and E.C. Pavlis (2001): 
On the possibility of measuring accurately the PPN parameters $\beta$ and $\gamma$ with laser-ranged satellites, 
preprint http://www.arxiv.org/abs/gr-qc/0103088, 
submitted to Class. and Quantum Grav.
\bibitem{}
Iorio, L. and E. C. Pavlis (2001): Tidal satellite perturbations
and the Lense-Thirring effect, J. of the Geodetic Soc. of Japan,
\textbf{47}, 1, 169-173.
\bibitem{}
Kaula, W. M. (1966): Theory of Satellite Geodesy,  Blaisdell
Publishing Company, 124p.
\bibitem{}
Lemoine, F. G.,  et al. (1998): The Development of the Joint NASA
GSFC and the National Imagery Mapping Agency (NIMA) Geopotential
Model EGM96, NASA/TP-1998-206861.
\bibitem{}
Lense, J. and  H. Thirring (1918): \"{U}ber den Einfluss der
Eigenrotation der Zentralk{\"{o}}rper auf die
 Bewegung der Planeten und Monde nach der Einsteinschen
Gravitationstheorie, Phys. Z., \textbf{19}, 156-163,  translated
by Mashhoon, B., F. W. Hehl and D. S. Theiss (1984): On the
Gravitational Effects of Rotating Masses: The Thirring-Lense
Papers, Gen. Rel. Grav., \textbf{16}, 711-750.
\bibitem{}
Montenbruck, O. and E. Gill (2000): Satellite Orbits: Models,
Methods, Applications, Springer, 383p.
\bibitem{} Pireaux, S., J.--P. Rozelot and S. Godier (2001):
Solar quadrupole moment and purely
relativistic gravitation contributions to Mercury's perihelion
advance, preprint http://www.arxiv.org/abs/astro-ph/0109032, submitted to Astrophysics and
Space Sciences.
\bibitem{}
Sengoku, A., M. K. Cheng and B. E. Schutz (1995): Anisotropic
reflection effect on satellite, Ajisai, J. of Geodesy,
\textbf{70}, 140-145.
\bibitem{}
Sengoku, A., M. K. Cheng, B. E. Schutz and H. Hashimoto (1997):
Earth-heating effect on Ajisai, J. of the  Soc. of Japan,
\textbf{42}(1), 15-27.
\bibitem{}
Shapiro, I. et al. (1972): Mercury' s Perihelion Advance:
Determination By Radar, Phys Rev. Lett., \textbf{28}(24),
1594-1597.
\bibitem{}
Shapiro, I. (1990): Proc. of the 12th International
 Conference on General Relativity and Gravitation,
1989, University of Colorado at Boulder,  Cambridge University
Press.
\bibitem{}
Sterne, T. E. (1960): An Introduction to Celestial Mechanics,
Interscience, 206p.
\bibitem{}
Will, C. M. (1993): Theory and Experiment in
Gravitational Physics, 2nd edition, Cambridge University Press, 380p.
\end{thebibliography}
\end{document}